\documentclass{aa}
\usepackage{aas_macros}
\usepackage{amsmath}
\usepackage{natbib}
\usepackage{txfonts}
\usepackage{graphicx}
\usepackage{acronym}
\usepackage{xspace}
\usepackage{datetime}

\newcommand{\kev}[1][]{#1 \ensuremath{\mathrm{keV}}\xspace}
\newcommand{\unit}[2][]{#1\:\ensuremath{\mathrm{#2}}}
\newcommand{\pten}[2]{\ensuremath{\mathrm{#1 \times 10^{#2}}}}
\newcommand{\dms}[3]{\ensuremath{#1\degr\,#2^{\prime}\,#3^{\prime\prime}}}
\newcommand{\hms}[3]{\ensuremath{#1^{\mathrm{h}}\,#2^{\mathrm{m}}\,#3^{\mathrm{s}}}}
\newcommand{\amin}[1][]{#1$^{\prime}$}
\newcommand{\asec}[1][]{#1$^{\prime\prime}$}

\newcommand{\src}[1]{\acl{#1}}

\newcommand{\xmm}{XMM-\textit{Newton}\xspace}
\newcommand{\Chandra}{\textit{Chandra}\xspace}
\newcommand{\pn}{EPIC-pn\xspace}
\newcommand{\mos}{EPIC-MOS\xspace}
\newcommand{\Mes}{M87\xspace}
\newcommand{\ang}{\,\AA\xspace}

\graphicspath{{img/}}

\acrodef{WFC}{Wide Field Camera}
\acrodef{EPIC}{European Photon Imaging Camera}
\acrodef{SAS}{the \xmm Science Analysis Software}
\acrodef{ARF}{Ancillary Response File}
\acrodef{RMF}{Response Matrix File}
\acrodef{DEM}{Differential Emission Measure}
\acrodef{CIE}[\textit{CIE}]{Collisionally Ionisation Equilibrium}
\acrodef{ICM}{Intra-Cluster Medium}
\acrodef{AGN}{Active Galactic Nucleus}
\acrodef{HRI}{High Resolution Instrument}
\acrodef{PSF}{Point Spread Function}
\acrodef{SNR}{Supernova Remnants}
\acrodef{SN}{supernovae}
\acrodef{RGS}{Reflection Grating Spectrometer}
\acrodef{OOT}{Out Of Time}
\acrodef{LHB}{Local Hot Bubble}
\acrodef{CXB}{Cosmic X-ray Background}
\acrodef{EPL}{Extra-galactic Power Law}
\acrodef{SN Ia}{Type Ia \ac{SN}}
\acrodef{SOC}{Science Operations Centre}
\acrodef{IMF}{Initial Mass Function}
\acrodef{AGB}{Asymptotic Giant Branch}

\acrodef{5044}{NGC 5044}
\acrodef{5813}{NGC 5813}

\title {The metal contents of two groups of galaxies.}

\abstract{The hot gas in clusters and groups of galaxies is continuously being enriched with metals from supernovae and stars. It is well established that the enrichment of the gas with elements from oxygen to iron is mainly caused by supernova explosions. The origins of nitrogen and carbon are still being debated. Possible candidates include massive, metal-rich stars, early generations of massive stars, intermediate or low mass stars  and \ac{AGB} stars. } 
{In this paper we accurately determine the metal abundances of the gas in the groups of galaxies \src{5044} and \src{5813}, and discuss the nature of the objects that create these metals. We mainly focus on carbon and nitrogen.}
{We use spatially-resolved high-resolution X-ray spectroscopy from \xmm. For the spectral fitting, multi-temperature hot gas models are used.} 
{The abundance ratios of carbon over oxygen and nitrogen over oxygen that we find are high compared to the ratios in the stars in the disk of our Galaxy. The oxygen and nitrogen abundances we derive are similar to what was found in earlier work on other giant ellipticals in comparable environments. We show that the iron abundances in both our sources have a gradient along the cross-dispersion direction of the \ac{RGS}.}
{We conclude that it is unlikely that the creation of nitrogen and carbon takes place in massive stars, which end their lives as core-collapse supernovae, enriching the medium with oxygen because oxygen should then also be enhanced. Therefore we favour low-and intermediate mass stars as sources of these elements. The abundances in the hot gas can best be explained by a 30--40\% contribution of type Ia supernovae based on the measured oxygen and iron abundances and under the assumption of a Salpeter \ac{IMF}.}

\author{ Y.G. Grange \inst{\ref{ins:SRON}} \and J. de Plaa \inst{\ref{ins:SRON}} \and J.S. Kaastra \inst{\ref{ins:SRON},\ref{ins:UU}}  \and N. Werner \inst{\ref{ins:Stan}} \and F. Verbunt \inst{\ref{ins:UU}, \ref{ins:SRON}} \and F. Paerels \inst{\ref{ins:Colum}} \and C.P. de Vries \inst{\ref{ins:SRON}}}
\institute{SRON Netherlands Institute for Space Research, Sorbonnelaan 2, 3584 CA Utrecht, The Netherlands \label{ins:SRON}
\and Astronomical Institute, Utrecht University, PO Box 80000, 3508 TA Utrecht, The Netherlands \label{ins:UU}
\and Kavli Institute for Particle Astrophysics and Cosmology, Stanford University, 382 Via Pueblo Mall, Stanford, CA94305-4060, USA\label{ins:Stan}
 \and Columbia Astrophysics Laboratory and Department of Astronomy, Columbia University, 550 West 120th Street, New York, NY10027, U.S. \label{ins:Colum}
}

\offprints{y.g.grange@sron.nl}

\date{Received 23 November 2010 / Accepted 30 March 2011}

\begin{document}
\maketitle
\acresetall
\section{Introduction} \label{sec:int}
Most of the galaxies in the Universe reside in groups \citep[e.g.][]{TUL87}. Groups of galaxies typically contain up to tens of galaxies. Due to the fact that their gravitational potential is relatively shallow, groups of galaxies are good laboratories to study non-gravitational effects \citep[see e.g. the review by][]{MUL04}. The gravitational compression and infall shocks, which heat up the gas, depend on the depth of gravitational potential, therefore groups of galaxies are cooler than the more massive clusters.
\par Since the temperature of groups is rather low (\unit[$\sim 10^{7}$]{K}), the emission lines of metals present in the medium are relatively strong. This yields an insight in the abundance distribution of metals in the intra-group medium. Since all the metals in the intra-group medium have been produced by stars, metal abundances in the intra-group medium are a tracer of the history of stellar evolution. 
\acused{EPIC}\acused{RGS}
\par X-ray spectroscopy of groups and clusters of galaxies is a powerful tool to measure the abundances in the hot gas. The broad spectral band of the \aclp{EPIC} (\acs{EPIC}-MOS, \citealt{TUR01} and \acs{EPIC}-pn, \citealt{STR01}) and the high spectral resolution of the \aclp{RGS} (\acs{RGS}, \citealt{HER01}) on board of \xmm allow us to determine the key abundances needed to constrain supernova models and ratios \citep[see e.g. the review by][]{BOH09}. While earlier work focussed on clusters, more recently, the metal contents of groups has been investigated in more detail \citep[e.g.][]{RAS07, WER09}. Measurements with the \ac{RGS} are the only way to constrain the carbon and nitrogen abundance in the hot gas.
\par The abundances observed in the intracluster medium are the result of the production of the elements in stars and the transport of the elements released to their current location.The metals with even atomic numbers from oxygen to up to iron and nickel are produced in supernovae: oxygen, neon and magnesium mainly in core-collapse supernovae, iron and nickel mainly in type Ia supernovae. Both types create significant amounts of silicon, sulfur, argon and calcium. The relative fraction of these elements is however unknown and highly model dependent \citep[see e.g.][]{IWA99,WOO02}. Carbon and nitrogen are produced by a variety of sources, the relative importance of which is still under debate \citep[for an overview of possible sources of several elements, see][and references therein]{ROM10}. Based on optical observations of stars in our galaxies, most authors conclude that superwinds of massive metal-poor stars enriched the gas with carbon early in the Galaxy disk's evolution. Some authors argue that at a later stage, a significant amount of carbon is created by low-mass stars \citep[e.g.][]{CES09}. Others \citep[e.g.][]{BEN06,MAT10} advocate the enrichment at later stages is mainly due to intermediate mass stars, during their \ac{AGB} phase. \citet{BEN06} conclude from the literature that it is likely that the enrichment with carbon is probably due to a complicated, and finely tuned set of objects.
Nitrogen is created by the conversion of carbon and oxygen during the CNO cycles of hydrogen burning in low and intermediate mass stars \citep{EDM78,CHI03} and metal-poor massive stars \citep{MAT86,MEY02}.
\par Several mechanisms were proposed that enrich the gas in groups and clusters of galaxies with the metals created inside the galaxies. An overview of these mechanisms is given by \citet{SCH08}. Simulations suggest that the gas is mainly enriched by galactic winds, powered by core-collapse supernovae at early stages of the evolution. During the life time of the group, the winds get more and more driven by type Ia supernovae. Other mechanisms include e.g. uplift of metals by the central \ac{AGN} and ram-pressure stripping of infalling galaxies. The relative importance of these processes is however yet unknown.
\par In this paper, we study two X-ray bright groups of galaxies, \object{NGC 5044} and \object{NGC 5813}. Both groups are named after the giant elliptical galaxy in their core. Multiple dwarf galaxies are found around the central elliptical galaxy of these groups. \src{5044} has previously been observed with several X-ray satellites (\textit{Einstein}: \citealt{FAB92}, \textit{ROSAT}: \citealt{DAV94}, \textit{ASCA}: \citealt{FUZ96}). The X-ray emission is sharply peaked in the central region and the temperature declines towards the core. Recently, the kinematic structure of \src{5044} was studied with \Chandra by \citet{DAV09}. \Chandra images show multiple filaments and cavities, which indicate activity of the central \ac{AGN} \citep{GAS09}. \xmm observations show a clear multi-temperature structure and \citet{TAM03}  put constraints on the metal abundances. Since then, the models and instrumental calibration have been improved and longer observations have been performed to measure the abundances with a higher precision. 
\par The group around \src{5813} is part of a larger gravitationally bound system which also includes the group around \object{NGC 5846}. \src{5813} has a kiloparsec size, kinematically decoupled core \citep{EFS82} caused by a history of minor or intermediate mergers long ago. It is classified by \citet{EMS07} as being a slow rotator, which they conclude is an extreme evolutionary end point reached in a deep potential well. In essence, this means that this group is extremely relaxed. \citet{RAN11} studied the shocks caused by the \ac{AGN} feedback in this group using data taken by \Chandra, and conclude that the outflow of the central \ac{AGN} is energetic enough to heat up the gas. 
\par We have obtained deep \xmm observations of these two groups. Exploiting the high spectral resolution and sensitivity of \xmm, we aim to accurately determine metal abundances in the hot gas. Especially the carbon and nitrogen abundances contain clues on the likely candidates for their production. In addition, the ratio between type Ia and core-collapse supernovae can be constrained using the oxygen and iron abundance. We use the measured abundances in the intra-group medium to test supernova yield models.
\par Throughout this paper, we use $H_{0}=\unit[70]{km\,s^{-1}\,Mpc^{-1}}$, $\Omega_{\mathrm{M}}=0.3$, and $\Omega_{\Lambda}=0.7$. At the redshift of \src{5044} ($z=0.009$), \amin[1] represents a distance of \unit[11.1]{kpc}. At the redshift of \src{5813} ($z=0.007$), \amin[1] represents a distance of \unit[7.92]{kpc}. The abundances are in units of the proto-Solar abundances \citep{LOD03}. Quoted statistical uncertainties are at a $1\sigma$ (68\%)  confidence level. Results obtained from the literature using different abundance scales (\citealt{AND89} and \citealt{GRE98}) or significance levels (e.g. 90\%) were converted to the units of this work before comparing them. 
\section{Data reduction and analysis}
All the data used for this research were obtained using the instruments on board of the \xmm satellite. A log of the observations is presented in Table \ref{tab:obslog}. The recent (2008 and 2009) data were obtained using the thin filter and for the archival data the medium filter was used. We use data from both the \ac{EPIC} and the \ac{RGS}. The \mos was in full frame mode during all observations. For the \pn, the extended full frame mode was turned on during the 2005 observation of \src{5813}, while for all others it was in the Full Frame mode. In total, \unit[153]{ks} of data are available for
\src{5044} and \unit[173]{ks} for NGC 5813. The reduction was done with \ac{SAS} version 9.0.0 with the calibration files created on 22 June 2009. 
\par In order to reduce contamination from soft-proton flares, we created a lightcurve for each instrument and observation. For \ac{EPIC}, the light curve was based on the data in the 10--12\kev energy band and for \ac{RGS} we used the data from CCD number 9, where hardly any emission from the group is expected. We binned the light curve in \unit[200]{s} intervals  for \ac{RGS} and \unit[100]{s} intervals for \ac{EPIC}. A Poissonian distribution was fitted to the count-rate histogram. We rejected all time bins for which the number of counts lies outside the interval $\mu \pm 2\sqrt{\mu}$, where $\mu$ is the fitted average of the Poissonian. The clean exposure times are listed in Table \ref{tab:obslog}. Since we cut off all data above $\mu + 2\sigma$ as well as all data under $\mu - 2\sigma$, we do not expect the flux to change due to our filtering. To check whether a less severe filtering would significantly improve the statistics, we fitted \ac{RGS} spectra for which we did not use any filtering at all. These fits are statistically comparable to the results of this work. In addition, we created a light curve from the events between 0.3 and 1 \kev to check whether our spectra are contaminated by charge-exchange emission from the interactions between the solar wind and gas in the ecliptic plane of our solar system and the Earth's atmosphere. This emission should show up as flares in the light curve under \kev[1]. The light curve does not show significant flares in this energy region.
\begin{figure}[t]
\centerline{
 \includegraphics[width=0.45\textwidth]{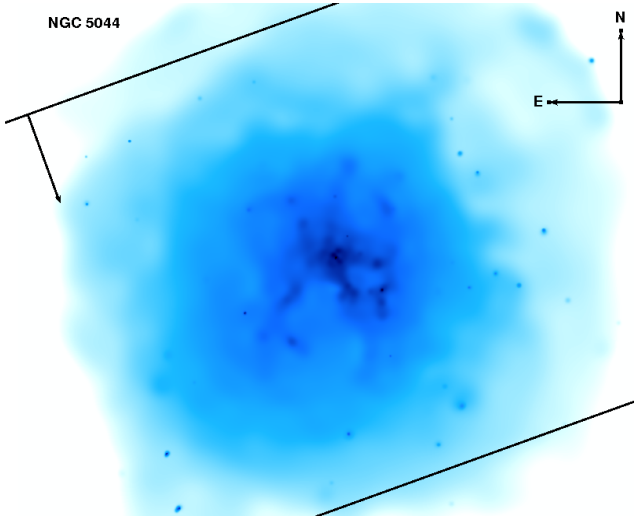}
}
\caption{Smoothed \Chandra image of \src{5044}. The edges of the \ac{RGS} for obsID 0037950101 are indicated with black lines. The black arrow has length \amin[1], and points in the positive cross-dispersion direction.}
\label{fig:Cha5044}
\end{figure}
\begin{figure}[t]
\centerline{
\includegraphics[width=0.45\textwidth]{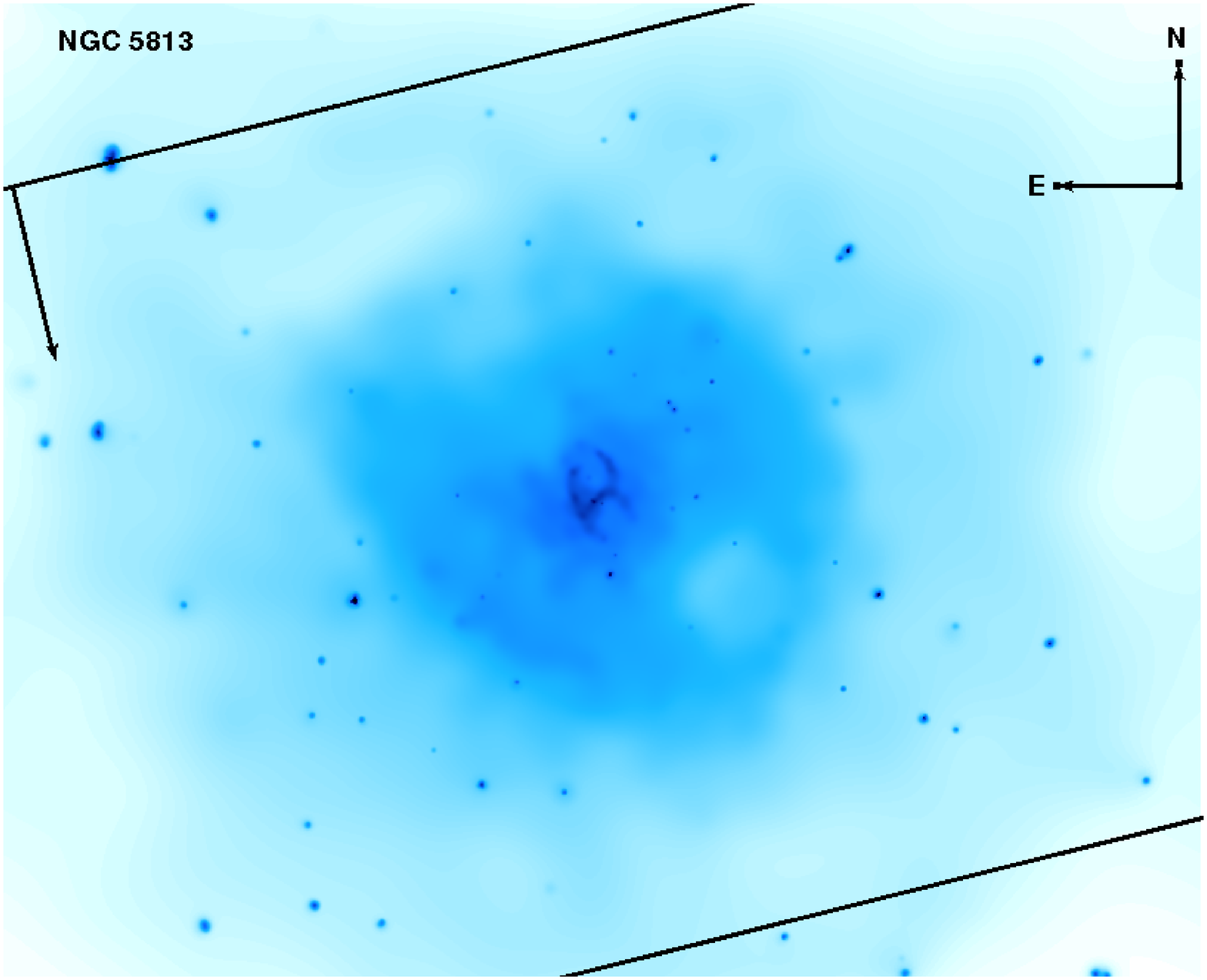}
}
\caption{Smoothed \Chandra image of \src{5813}. The edges of the \ac{RGS} for obsID 0554680201 are indicated with black lines. The black arrow has length \amin[1], and points in the positive cross-dispersion direction.}
\label{fig:Cha5813}
\end{figure}

\begin{table}[t]
\caption{Log of the observations used.}  \label{tab:obslog}
\centerline{
\scalebox{0.85}{
\begin{tabular}{llllll}
\hline
\hline
       & Exposure  & Exposure   & Exposure & & Position\\
obsID  & MOS   & pn     & RGS    & Start date &   angle\\
       & (ks)   & (ks)  & (ks)       & (Y-M-D) &(degrees)\tablefootmark{a}\\
\hline
\multicolumn{6}{c}{NGC 5044} \\
\hline
0037950101 & 21  & 12 & 17 & 2001-01-12 & 110 \\
0554680101 & 104 & 86 & 95 & 2008-12-27 & 114 \\
\hline
Total exposure & 125 & 98 & 112 & &\\
\hline
\multicolumn{6}{c}{NGC 5813} \\
\hline
0302460101 & 30 & 22 & 25 & 2005-07-23 & 291 \\
0554680201 & 60 & 44 & 51 & 2009-02-11 & 104 \\
0554680301 & 60 & 43 & 50 & 2009-02-17 & 102 \\
\hline
Total exposure & 150 & 109 &  126 & & \\
\hline
\end{tabular}
}
}
\tablefoot{
Quoted exposures are after filtering the soft-proton background.\\
\tablefoottext{a}{The zeropoint of the position angle is the celestial north. The positive position-angle direction is towards the east.}
}\end{table}
\par The \ac{RGS} is a slitless instrument. This means that the zeropoint of the wavelength calibration is dependent on the position of the source in the detector field of view. We gauge our wavelength scale of \ac{RGS} on the peak of the emission in the X-ray image. We also use this position as the source centre for the \ac{EPIC} extraction region. For \src{5044}, we take $\alpha$=\hms{13}{15}{24}, $\delta$=\dms{-16}{23}{08} and for \src{5813}, these coordinates are $\alpha$=\hms{15}{01}{11}, $\delta$=\dms{01}{42}{06}. These coordinates are within \asec[2] from the optical position of the central galaxy.
\subsection{Reduction of the RGS data}
The RGS data were reduced using the standard pipeline processing of the \ac{SAS}. We subtract from the total spectrum a model background spectrum which is created by the standard \ac{RGS} pipeline. This method consists of selecting a template background file, based on the count rate in CCD 9 of the RGS. We do not explicitly model the cosmic X-ray background in \ac{RGS} because any diffuse line source would be smeared out into a broad, nearly flat continuum-like component. We included 100\% of the cross-dispersion direction in the spectrum because the extent of the sources is larger than the width of the \ac{RGS}. All other options were left to their default values. To make one-dimensional spatial profiles of the parameters that we measured using \ac{RGS}, we also created spectra based on five narrow regions in the cross-dispersion direction of the RGS, within \amin[2] from the core. During one of the observations of \src{5813} (obsID 0302460101) the position angle differs by $\sim$180\degr\ from the other two observations. We fitted the spectra from the cross-dispersion strips of this observation simultaneously with the spectra from the opposing regions of the other two observations. In the other cases, we assumed that the differences in roll angle were not large enough to introduce systematic effects. The relative position of the sources to the \ac{RGS} detectors and the positive cross-dispersion directions are shown in Figs. \ref{fig:Cha5044} and \ref{fig:Cha5813}. 
\par Since, as mentioned before, \ac{RGS} is a slitless instrument, all photons within the field of 
view end up in the final spectrum. This causes a broadening of the 
spectral lines. If a photon originates from an angle $\Delta \theta$ (in arc 
minutes) projected along the dispersion direction, it will be registered as 
having a wavelength which is shifted by $\Delta \lambda$ (in \ang) with 
respect to its original wavelength. The \xmm Users Handbook gives the relation
\begin{equation}
\Delta \lambda = 0.138 \Delta \theta / m \text{.} \label{eq:dc2wv}
\end{equation}
Here, $m$ is the (absolute value of the) spectral order. We extracted the brightness profile of the source in the dispersion direction from the \mos image and used this to account for the broadening following the method developed by \citet{TAM04}. We use the spectral data between 7\ang and 36\ang. This upper limit is chosen so that we include the \ion{C}{vi} line at 33.7\ang. 

\subsection{EPIC data reduction}
Because we want to compare our \ac{EPIC} results with \ac{RGS}, we derive \ac{EPIC} spectra from a circle with radius \amin[3] around the centre. The diameter corresponding to this radius is comparable to the cross-dispersion width of the \ac{RGS} (\amin[5]). 
We identified point sources, based on visual selection, and excised a region with a radius of \asec[15] around them. In total, we excise\textit{d} one point source from the \src{5044} observation. Two point sources were removed which lie on the edge of the \amin[3] circle. From the field around \src{5813}, three point sources were identified and removed. Based on the spectrum, we created files containing the effective area and response matrix using the default SAS-tasks. To correct for vignetting effects, we created these files using a detector map of the source. Furthermore, we corrected the pn observations for \acf{OOT} events. We ignore all events corresponding to energies higher than \kev[10]. For \mos, we ignore the events below \kev[0.4]. We also ignore all events below \kev[1.5] in the \pn data because these data are affected by a gain shift, leading to large-scale systematic residuals below this energy. This effect seems to be present in both sources. 
\subsubsection{\ac{EPIC} background}
To correctly fit the \ac{EPIC} spectra, we need to subtract or model the background. \citet{SNO04} used a method to subtract the \mos instrumental background by scaling closed filter observations, as created by the \xmm \ac{SOC}. Other authors  \citep{ARN01, REA03, DEL04, NEV05} developed and used comparable methods. The methods mentioned are highly dependent on the region of the CCD outside the field of view. In the pn, this region is  very small which makes the method sensitive to statistical fluctuations. Since we want to simultaneously fit the three instruments, we need to use a unified background subtraction method. Our method involves modelling the background by use of additional spectral models \citep[see e.g.][]{LEC08a,PLA10}, as follows. We model both instrumental and Cosmic X-ray background.
\par For the instrumental background, we used the closed filter observations, provided by the \xmm \ac{SOC}. We extracted a spectrum from these closed filter files, using the same procedure as for the actual data. We modeled this instrumental background using a broken power law, which describes the hard particle spectrum, and a number of delta functions that represent the fluorescence lines caused by the interaction of hard particles with the instrument itself. We list the fluorescence lines that we take into account in Table \ref{tab:EPIC-lines}. Since high-energy particles are not focussed by the telescope in the same way as X-ray photons, we do not fold these models through the \ac{ARF}, which describes the effective area of the instrument. The instrument degrades with time, leading to an increase of the instrumental background. In addition, the instrumental powerlaw does not vary over the field of view, while the strength of the instrumental lines does depend on the position on the detector. During the fits to the core spectra, we fitted the instrumental background for each observation by allowing the normalisations of the instrumental background components to vary within a factor of 4 of the best-fit value, derived from the closed filter spectrum. We also left free the power-law break strength. We assumed that the two observations of \src{5813} on February 2009 had the same instrumental background.
\begin{table}[t]
\caption{Instrumental lines modeled for \mos and \pn.} 
\label{tab:EPIC-lines}
\centerline{
\begin{tabular}{llll}
Element & Energy & MOS & pn \\
& (\kev) & & \\
\hline
\hline
Al K$\alpha$ & 1.486 & + & + \\
Al K$\beta$ & 1.557  & + & + \\
Si K$\alpha$ & 1.740 & + & $-$ \\
Si K$\beta$ & 1.835  & + & $-$ \\
Ti K$\alpha$ & 4.51  & $-$ & + \\
Cr K$\alpha$ & 5.41  & + & + \\
Mn K$\alpha$ & 5.89  & + & $-$ \\
Fe K$\alpha$ & 6.40  & + & $-$ \\
Ni K$\alpha$ & 7.47  & $-$ & + \\
Cu K$\alpha$ & 8.04  & $-$ & + \\
Cu K$\beta$ & 8.90   & $-$ & + \\
Zn K$\alpha$ & 8.63  & + & + \\
Zn K$\beta$ & 9.57   & $-$ & + \\
Au L$\alpha$ & 9.72  & + & + \\
\hline
\end{tabular}
}
\tablefoot{Lines from \citet{PLA10} and \citet{LEC08a}. A ``+'' indicates the line is present in the background model for the instrument. A ``$-$'' means the line is not part of the background model for the given instrument.}
\end{table}
\par The second group of components models the emission by the astrophysical back- and foreground. 
These X-rays are focussed by the mirror in the same way as source photons, and therefore these components should be folded through the \ac{ARF}. At high energies ($>2\kev$), the background is dominated by the \ac{CXB} of unresolved point sources. This component is modeled using a power-law, the \ac{EPL}. We fix the photon index $\Gamma$ to 1.41 and the \kev[2--10] flux to \unit[\pten{2.24}{-14}]{Wm^{-2}\deg^{-2}}  \citep{DEL04}. 
\par At lower energies ($<2\:\kev$), foreground emission from the \ac{LHB} dominates the background spectrum. 
Contrary to the other X-ray background components, this one is not absorbed by the neutral gas of our Galaxy. Furthermore, the diffuse halo or disk of the Galaxy also contributes to the foreground at these low energies \citep{KUN01}. We modeled these two components using a \ac{CIE} plasma model. Since the temperatures of these components vary across the sky, we have to fit the temperatures locally. Therefore, we extracted a spectrum from an annulus between \amin[9] and \amin[12] centred on the emission peak of the group. In this annulus the group emission is low. In addition to the background components, we added an extra \ac{CIE} component to model any residual group emission. The best-fit values of the relevant parameters are listed in Table \ref{tab:bgcomp}. We keep the parameters of the X-ray background model components frozen to the best-fit values while fitting the spectra in the core. The normalisations of the components are scaled to the size of the extraction region. 

\begin{table}[t]
\caption{Relevant parameters of the fitted background components.} \label{tab:bgcomp}
\centerline{
\begin{tabular}{llll}
\hline
\hline
Component & Parameter & \src{5044} & \src{5813} \\
\hline
\ac{EPL} & $\Gamma$\tablefootmark{a} & 1.41 & 1.41\\
    & Flux\tablefootmark{a,b}(\kev[2--10]) & 2.24 & 2.24 \\
    & Flux\tablefootmark{a,b,c} (\kev[0.1--2.5]) & 0.96 & 1.02 \\
\ac{LHB} & $kT$ (\kev) & 0.11 & 0.07 \\
    & Flux\tablefootmark{b} (\kev[2--10]) & \pten{1.57}{-7} &  \pten{2.05}{-12}\\
    & Flux\tablefootmark{b} (\kev[0.1--2.5])& 3.61 &  4.48\\
halo & $kT$ (\kev) &  0.29 & 0.19\\
    & Flux\tablefootmark{b} (\kev[2--10]) & \pten{1.84}{-3} & \pten{1.23}{-4} \\
    & Flux\tablefootmark{b} (\kev[0.1--2.5])& 0.78 & 0.90 \\
\hline
\end{tabular}
}
\tablefoot{
\tablefoottext{a}{Fixed during fit}
\tablefoottext{b}{In units of \unit[\pten{}{-14}]{Wm^{-2}\deg^{-2}}}
\tablefoottext{c}{Values differ because of the difference in galactic \ion{H}{i} values.}
}
\end{table}

\subsection{Spectral models} \label{subsec:mod}
We fit our spectra using the SPEX package \citep{KAA96}. Since clusters and groups are not isothermal, we use multi-temperature models. We fit the spectra using two different empirical parametrisations of a \ac{DEM} model. The emission measure $Y=n_\mathrm{e}n_\mathrm{h}\mathrm{d}V$ with $n_\mathrm{e}$ and $n_\mathrm{h}$ the electron and proton densities and $V$ the volume of the source.
\par The first parametrisation is the \textit{gdem} model, in which the differential emission measure $\mathrm{d}Y/\mathrm{d}T$ as a function of temperature is distributed as \citep[e.g.][]{PLA06}
\begin{equation}
\frac{\mathrm{d}Y}{\mathrm{d}T}=\frac{Y_{0}}{\sigma_{\mathrm{T}}\sqrt{2\pi}}\mathrm{e}^{-(x-x_{0})^{2}/2\sigma_{\mathrm{T}}^{2}} \text{.}
\end{equation}
In this equation, $x=\log(T)$ and $x_{0}=\log(T_{0})$. 
\par The second \ac{DEM} model we fit is the \textit{wdem} model \citep{KAA04_art}. The model consists of a number of thermal components distributed as a truncated power law. The model is characterised by:
\begin{equation}
\frac{\mathrm{d}Y}{\mathrm{d}T} = \left\{ \begin{array}{ll} 
cT^{1/\alpha} & \hspace{1.0cm} \beta T_{\mathrm{max}} \le T < T_{\mathrm{max}} \\
0 & \hspace{1.0cm} T > T_{\mathrm{max}} \lor T < \beta T_{\mathrm{max}} \\
\end{array} \right.\text{.}
\label{eq:wdem_dydt}
\end{equation}
For $\alpha\rightarrow 0$, the model becomes isothermal with temperature $kT_{\mathrm{max}}$. 
\par In order to compare the multi-temperature models, we calculate the emission-weighted temperature for \textit{wdem} \citep[see][for the derivation]{PLA06}, 
\begin{equation}
kT_{\mathrm{mean}} = \frac{(1 + 1/\alpha)}{(2 + 1/\alpha)} \frac{(1 - \beta^{1/\alpha + 2})}{(1 - \beta^{1/\alpha + 1})} kT_{\mathrm{max}} \text{.}
\label{eq:wdem_mean}
\end{equation}
For \textit{gdem} the peak temperature is fitted.
\par The hot gas models allow us to determine the abundances of metals in the intra-group medium. Since the neon and magnesium abundance could not be constrained using \ac{EPIC}, we fixed the ratio between these and iron to the values found in RGS. For the Galactic foreground absorption due to neutral gas, we used the \ion{H}{i} column density from the Leiden/Argentine/Bonn survey \citep{KAL05}. For \src{5044} we use a column density of \unit[\pten{4.87}{24}]{m^{-2}} and for \src{5813} the value is \unit[\pten{4.37}{24}]{m^{-2}}. 

\section{Results}\label{sec:res}
We fitted the spectra of \src{5044} and \src{5813} using the models described in section \ref{subsec:mod}. In order to easily compare the abundances found to other groups, we quote abundance ratios for most elements. We chose to quote the abundances relative to iron, because iron has the smallest statistical uncertainty of all measured abundances. 
\subsection{\src{5044}}
\begin{figure}[t]
\centerline{
 \includegraphics[angle=-90,width=0.45\textwidth]{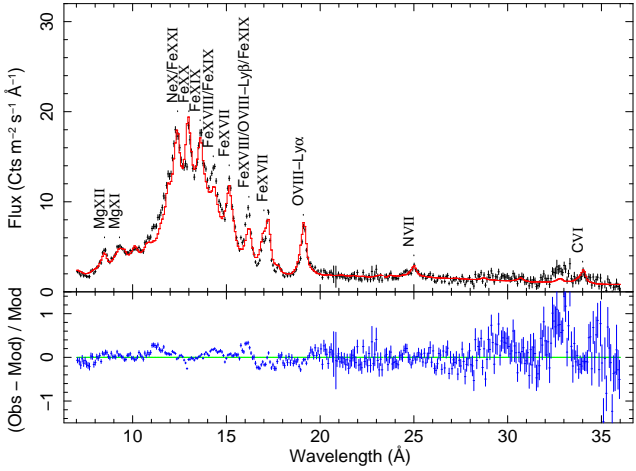}
}
\caption{Fluxed, background-subtracted \ac{RGS} spectrum of \src{5044}. The (red) line through the data points is the best fit. The most important spectral lines are indicated in the plot.} \label{fig:5044-rgspec}
\end{figure}

In Fig. \ref{fig:5044-rgspec}, we show the \ac{RGS} spectrum of \src{5044}. The best-fit values are listed in Table \ref{tab:res5044}. The spectrum shows a wealth of lines that can be used to constrain the abundances and multi-temperature structure of the gas. Most lines are fitted well, however both the \ion{Fe}{xviii} and \ion{O}{viii}-Ly$\beta$ line blend around 16\ang and the blend of \ion{Fe}{xviii} and \ion{Fe}{xix} near 14\ang are not fitted optimally. Near 11\ang the model is also underestimating the data.
\par Table \ref{tab:res5044} shows that this source has a significant temperature structure to which the \textit{gdem} model fits slightly better than the \textit{wdem} model. The derived abundance ratios, as reported in Table \ref{tab:res5044}, are consistent. The absolute iron abundances derived from these models differ by about $3\sigma$. The mean temperature is consistent between the two models.
\par The background for \ac{RGS} at long wavelength is high for extended sources, and hence some systematic residuals remain near the \ion{C}{vi} line at $\sim34$\ang.

\begin{figure}[t]
\centerline{
 \includegraphics[angle=-90,width=0.45\textwidth]{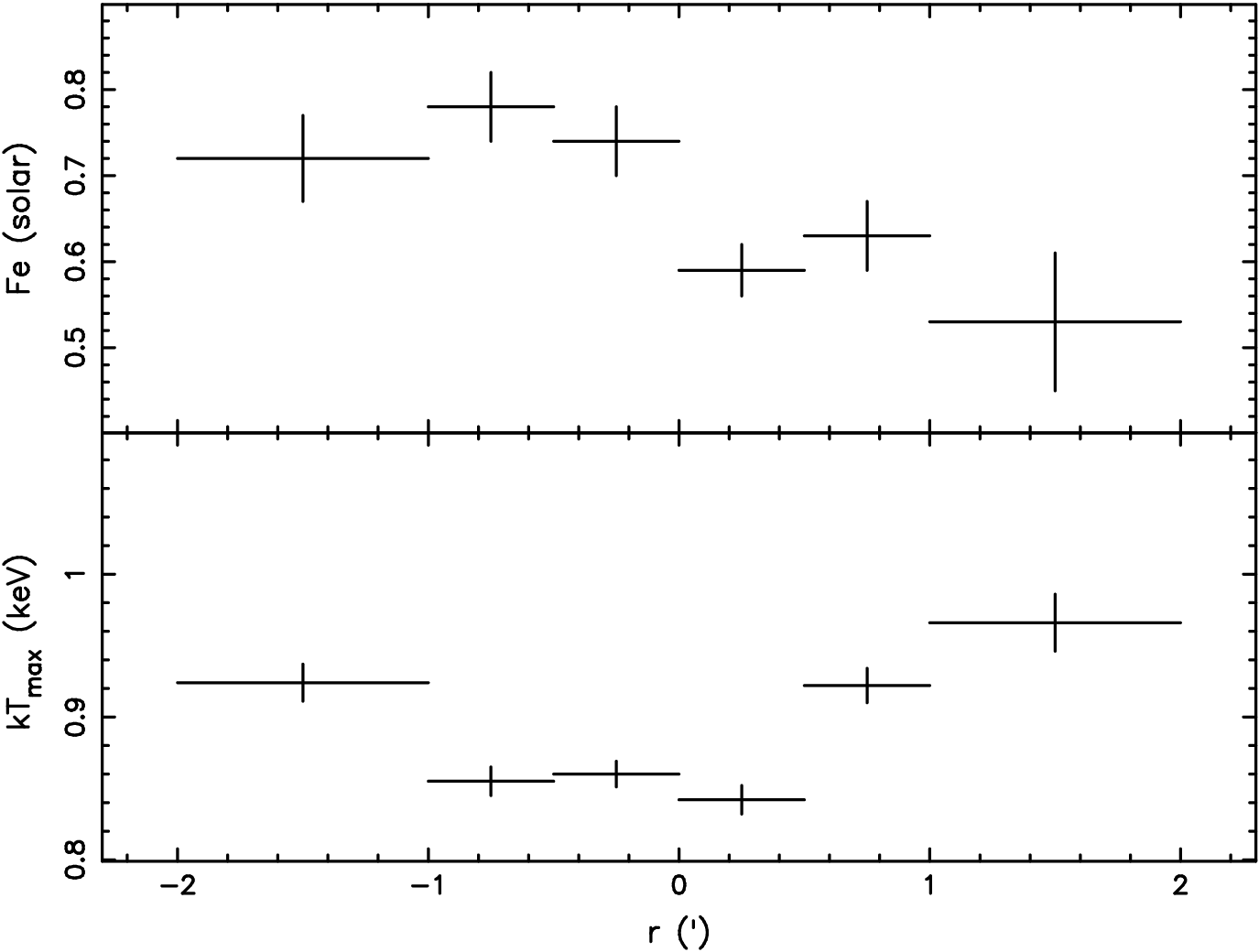}
}
\caption{Iron and temperature profiles in the cross-dispersion direction based the fits to the \ac{RGS} data of \src{5044} using the \textit{gdem} model.} \label{fig:5044-rgsrad}
\end{figure}
\par
\begin{figure}[t]
\centerline{
 \includegraphics[angle=-90,width=0.45\textwidth]{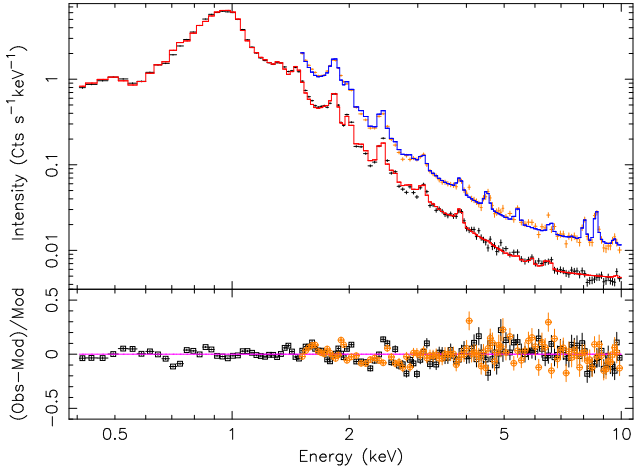}  
}
\caption{EPIC spectrum of \src{5044}. The top panel shows the data points with the model as a histogram through them. Upper data set: \pn, bottom data set: \mos. The bottom plot shows the fitting residuals. Squares: \pn, circles:\mos.} \label{fig:5044-mpspec}
\end{figure}

 Fig. \ref{fig:5044-rgsrad} shows the profiles of the \textit{gdem} peak temperature and iron abundance as measured along the cross-dispersion direction of \ac{RGS} (see Fig. \ref{fig:Cha5044} for the physical direction of the cross-dispersion direction). The temperature shows an increase from the centre of the galaxy towards the outer parts, as is normal in the cooling cores of groups and clusters of galaxies. The iron abundance shows a hint of a linear decrease over the cross-dispersion direction. Fitting a linear model to the data yields a normalisation of $0.663\pm0.016$ solar at $r=0$ and the slope of the line is $-0.08\pm0.02$ solar per arc minute.
\begin{figure}[t]
\centerline{
 \includegraphics[angle=-90,width=0.45\textwidth]{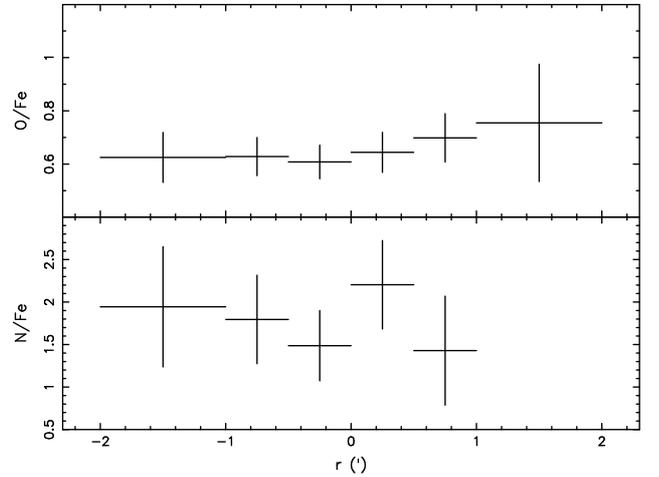}
}
\caption{O/Fe and N/Fe profiles (in proto-Solar units by \citealt{LOD03}) based on the fits to the \ac{RGS} data of \src{5044} using the \textit{gdem} model. The nitrogen abundance could not be constrained in the rightmost bin and is therefore not present in the plot.} \label{fig:5044-NandO}
\end{figure}
The profiles of the relative abundances of oxygen and nitrogen are presented in Fig. \ref{fig:5044-NandO}. The oxygen over iron abundance shows no significant variation over the cross-dispersion direction. The nitrogen abundance is poorly constrained.
The \ac{EPIC} spectrum, (Fig. \ref{fig:5044-mpspec}) shows systematic residuals above 2\kev, where the background is becoming more important possibly to a low contribution of quiescent soft-protons. The residuals also show some systematic effects in the iron-L complex at energies below \kev[2]. 
\par The iron abundance derived using the \ac{EPIC} data is significantly higher than the one derived from \ac{RGS}. For the oxygen over iron ratio, the value derived from \ac{EPIC} is lower than the \ac{RGS} value. The abundance ratios of silicon, sulfur, argon, calcium and nickel over iron are high. The \ac{EPIC} fits give a temperature structure with a higher peak temperature and a lower emission measure than for those of the \ac{RGS}. For the \ac{EPIC} spectra, the derived temperature structure is broader than what we find for the \ac{RGS} spectra.

\begin{table}
\caption{Fitted values to the \src{5044} spectra.} \label{tab:res5044}
\centerline{
\begin{tabular}{llll}
\hline
Parameter & RGS \textit{wdem} & RGS \textit{gdem} & EPIC \textit{gdem} \\
\hline \hline
Y (\unit[$10^{71}$]{m^{-3}}) & 3.02(6)& 2.95(5) & 2.555(10) \\
$T_{\text{peak}}$ (keV) & 1.020(10) &  0.910(5) & 0.9585(14)\\
$T_{\text{mean}}$ (keV) & 0.887(10) & - & - \\
$\sigma_{\mathrm{T}}$ & - & 0.100(3) &  0.1464(12) \\
$1/\alpha$ & 0.178(9) & - & - \\
$\beta$\tablefootmark{a} & 0.4 & - & - \\
C / Fe & 1.5(3) & 1.5(3) & - \\
N / Fe & 1.4(2) & 1.5(2) & - \\
O / Fe &  0.61(3) & 0.60(3) & 0.472(17) \\
Ne / Fe & 0.35(12) & 0.37(10) & - \\
Mg / Fe & 0.80(6)  & 0.73(5) &  -\\
Si / Fe & - &- &  0.808(12)\\
S  / Fe & - &- &  0.871(18) \\
Ar / Fe & - &- &  0.789(13) \\
Ca / Fe & - &- &  1.84(13) \\
Fe &  0.572(14) & 0.616(16) & 0.782(6)\\
Ni / Fe & - &- &   1.60(5) \\
C-stat / d.o.f. &  2591 / 1332 & 2418 / 1332 & 3781 / 1775 \\
\hline
\end{tabular}
}
 \tablefoot{Abundance ratios are given in proto-Solar units by \citet{LOD03}. Numbers between parentheses are the $1\sigma$ errors on the same number of least significant digits. \tablefoottext{a}{Kept fixed during the fit.}}
\end{table}
\subsection{\src{5813}}
\begin{figure}[t]
\centerline{
 \includegraphics[angle=-90,width=0.45\textwidth]{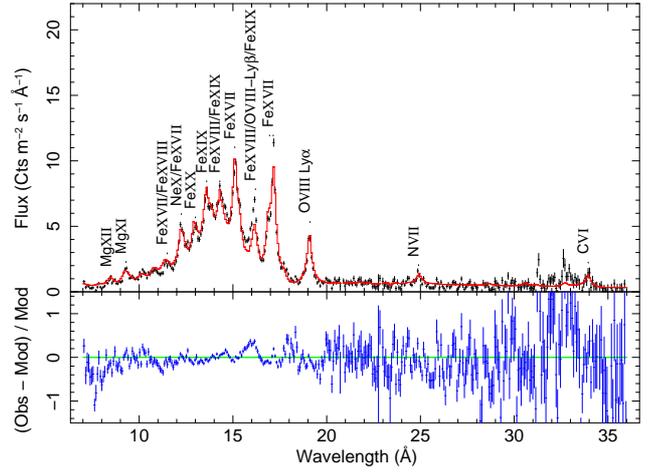}
}
\caption{Fluxed \ac{RGS} spectrum of \src{5813}. The (red) line through the data points is the best fit. The most important spectral lines are indicated in the plot.} \label{fig:5813-rgspec}
\end{figure}

We present the \ac{RGS} spectrum of \src{5813} in Fig. \ref{fig:5813-rgspec}. The model is underestimating the line flux of the oxygen and iron line blend near 16\ang, which is also the case for \src{5044}. Contrary to the fit for \src{5044}, the 14\ang blend is well fitted by the model.
\par The values and accuracies of the carbon and nitrogen abundances are well defined and comparable to what we find for \src{5044}. Also for this source, the background at $\sim34$\ang introduces extra uncertainty in the measured carbon abundance. 
\par The fits to the \ac{RGS} spectrum of \src{5813} show no significant difference between \textit{wdem} and \textit{gdem}. Both models converge to the single-temperature limit with essentially the same values for all parameters and the same statistical significance. 
\begin{figure}[t]
\centerline{
 \includegraphics[angle=-90,width=0.45\textwidth]{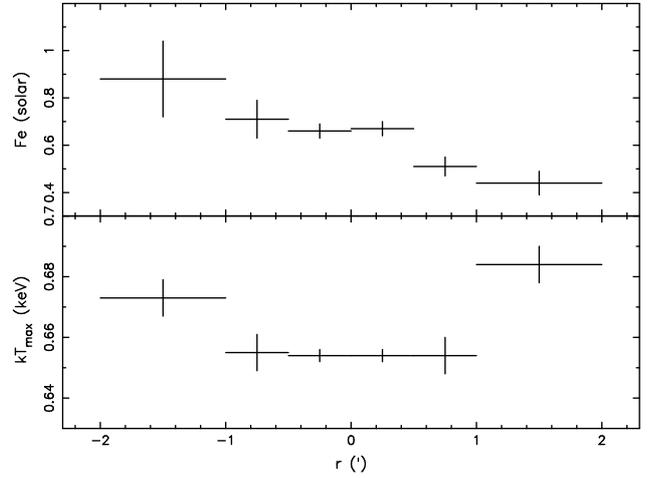}
}
\caption{Iron and temperature profiles in the cross-dispersion direction based on the fits to the \ac{RGS} data of \src{5813} using the \textit{gdem} model.}
\label{fig:5813-rgsrad}
\end{figure}
\par The \ac{RGS} profiles along the cross-dispersion direction of the \textit{gdem} peak temperature and the iron abundance that we obtained from the \ac{RGS} data are shown in Fig. \ref{fig:5813-rgsrad}. For the physical meaning of the cross-dispersion direction, see Fig. \ref{fig:Cha5813}. A decrease towards the centre of the temperature is observed, like in cool-core groups. However, the central temperature gradient is shallower than for \src{5044}. The iron abundance appears to be asymmetrically distributed around the core. A linear fit to this data set yields a normalisation of $0.650\pm0.018$ solar at $r=0$ and a slope of $-0.13\pm0.03$ solar per arc minute.
\begin{figure}[t]
\centerline{
 \includegraphics[angle=-90,width=0.45\textwidth]{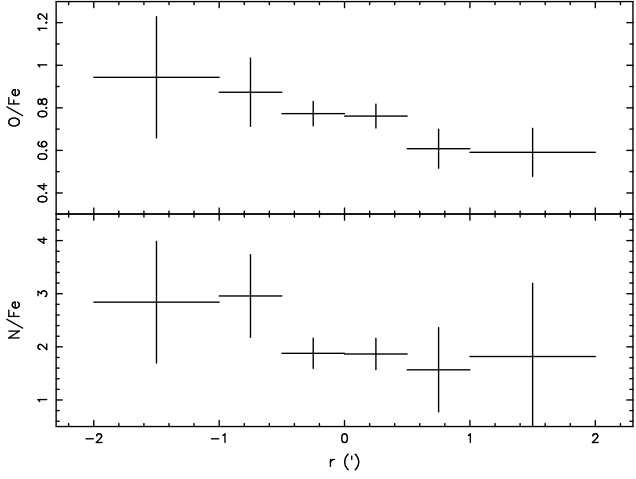}
}
\caption{O/Fe and N/Fe profiles (in proto-Solar units by \citealt{LOD03}) based the fits to the \ac{RGS} data of \src{5813} using the \textit{gdem} model.} 
\label{fig:5813-NandO}
\end{figure}
In Fig. \ref{fig:5813-NandO}, the radial profiles of oxygen and nitrogen over iron abundances are shown. The O/Fe abundance ratio is flat, with a hint of a decrease from the negative to the positive cross-dispersion direction.
\begin{figure}[t]
\centerline{
 \includegraphics[angle=-90,width=0.45\textwidth]{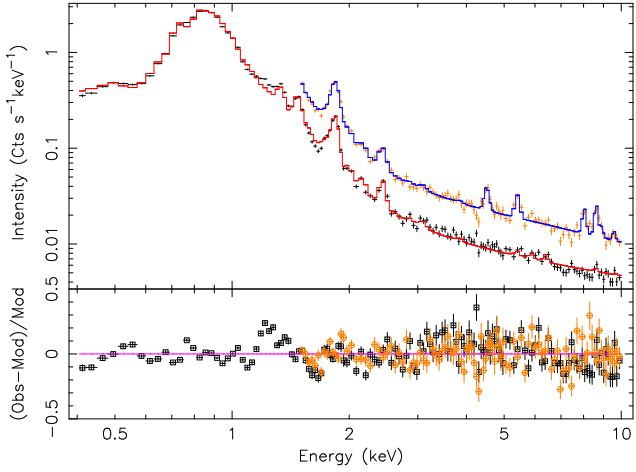}
}
\caption{EPIC spectrum of \src{5813}. The top panel shows the data points with the model as a histogram through them. Upper data set: \pn, bottom data set: \mos. The bottom plot shows the fitting residuals. Squares: \pn, circles: \mos.}
\label{fig:5813-mpspec}
\end{figure}
\par The \ac{EPIC} spectra of \src{5813} are shown in Fig. \ref{fig:5813-mpspec}. In the iron-L complex, around 1.2\kev some residuals remain which do not show up as strong in the spectra of \src{5044}. Between \kev[3] and \kev[5], the systematic residuals are similar to the same spectral regions in \src{5044}. 
\par As can be seen in Table \ref{tab:res5813}, the iron abundance in \ac{EPIC} is 16\% higher than the value deduced from \ac{RGS}. The emission measure is lower by 24\%, while the O/Fe ratio is 20\% lower. Contrary to \ac{RGS}, \ac{EPIC} shows a significant temperature structure with the \textit{gdem} model. Silicon, sulfur, argon and calcium to iron abundance ratios are high. The nickel to iron abundance is low. 
\begin{table}
\caption{Fitted values to the \src{5813} spectra.} \label{tab:res5813}
\centerline{
\begin{tabular}{llll}
\hline
Parameter & RGS \textit{wdem} & RGS \textit{gdem} & EPIC \textit{gdem} \\
\hline \hline
Y (\unit[$10^{70}$]{m^{-3}}) &  5.45(17) &  5.45(16) & 4.44(4) \\
$T_{\text{peak}}$\tablefootmark{a} (keV) & 0.67(3) & 0.665(2) & 0.6713(13) \\
$\sigma_{\mathrm{T}}$   & - & $<$ 0.012\tablefootmark{b} & 0.057(4) \\
$1/\alpha$ \tablefootmark{b}  & $<$ 0.08 & - & - \\
$\beta$ \tablefootmark{c} & 0.65 & - & - \\ 
C / Fe & 1.7(4)  &  1.7(4) & - \\ 
N / Fe & 1.8(3) &   1.8(3) & - \\
O / Fe & 0.72(5) &  0.73(4) & 0.61(3) \\
Ne / Fe & 0.58(10) &  0.58(10) & -\\  
Mg / Fe & 0.85(8) & 0.86(6)  & -\\
Si / Fe & - &- &  0.99(3)\\
S  / Fe & - &- &  1.09(5)\\
Ar / Fe & - &- &  1.0(2) \\
Ca / Fe & - &- & 1.9(5) \\
Fe &  0.538(19) & 0.538(17) & 0.588(8)\\ 
Ni / Fe & - &- &  0.54(9) \\
C-stat / d.o.f. & 3864 / 2667 &  3864 / 2667 & 3601 / 1775\\
\hline
\end{tabular}
}
 \tablefoot{Abundance ratios are given in proto-Solar units by \citet{LOD03}. Numbers between parentheses are the $1\sigma$ errors on the same number of least significant digits.  
\tablefoottext{a}{Since 1/$\alpha=0$, the \textit{wdem} mean and peak temperatures are equal.
\tablefoottext{b}{$2\sigma$ upper limit.}
 \tablefoottext{c}{Kept fixed during the fit.}}
 }
\end{table}
\section{Discussion}
We accurately determined the metal abundances of the hot gas in two groups of galaxies. In the following sections we focus the discussion on four main points. First, the derivation of the temperature structure, including the fitting of the iron-L complex and its effects on the reported abundance ratios. Second, we show how our abundance ratios compare to what was earlier reported in the literature. After this the supernova models and numbers using the abundances found in the gas are discussed. Finally we address the stellar origins of the carbon, nitrogen and oxygen abundances.

\subsection{Temperature structure}
\par In general, multi-temperature models fit the spectra of both groups well. In the \ac{EPIC} spectrum of \src{5813} we find some systematic effects in the residuals near 1.2\kev. \citet{BRI00} encountered the same feature in Capella. They solve this problem by adding lines caused by transitions of \ion{Fe}{xvii}, \ion{Fe}{xviii} and \ion{Fe}{xix} to the model. The abundances of Capella significantly increased when doing this. We recalculated the best fitted model using an experimental version of the line list that is being implemented in SPEX\footnote{Courtesy A.J.J. Raassen}, which contains many more transitions, including the ones probably causing the feature. Doing this did improve the fit near 1.2\kev. To date it is however not possible yet to use the new list for fitting yet.
\par In \ac{RGS}, we find that some iron lines in the iron-L complex are not well fitted in both groups. In Table \ref{tab:Feblend}, we list the ten strongest lines between 15.3\ang and 16.7\ang in the fitted models to our sources. The eight brightest lines are the same for both sources but their order in the list is slightly different. Most of these strongest lines are due to the \ion{Fe}{xviii} ion. 
\begin{table}
\caption{The ten strongest modeled lines around the 16\ang blend, using the fitted model to the sources.} \label{tab:Feblend}
\begin{center}
\begin{tabular}{lll}
\hline
Ion & Wavelength & Emission \\
& \AA & $\pten{}{47} \text{photons}\, s^{-1}$\\
\hline
\hline
\multicolumn{3}{c}{\src{5044}} \\
\hline
\ion{Fe}{xix}   & 15,358 & 11.7 \\ 
\ion{Fe}{xix}   & 15.364 & 16.1 \\ 
\ion{Fe}{xviii} & 15.831 & 24.3 \\
\ion{Fe}{xviii} & 16.165 & 26.1 \\
\ion{Fe}{xix}   & 16.282 & 26.4 \\
\ion{Fe}{xviii} & 15.628 & 38.9 \\
\ion{Fe}{xviii} & 16.002 & 41.0 \\
\ion{O}{viii}   & 16.003 & 62.1 \\
\ion{Fe}{xix}   & 16.106 & 87.2 \\
\ion{Fe}{xviii} & 16.078 & 120.9 \\
\hline
\multicolumn{3}{c}{\src{5813}} \\
\hline
\ion{Fe}{xix}   & 16.282 & 7.7 \\
\ion{Fe}{xviii} & 15,874 & 7.9 \\ 
\ion{Fe}{xvii}  & 15.456 & 9.9 \\ 
\ion{Fe}{xviii} & 15.831 & 16.9 \\
\ion{Fe}{xviii} & 16.165 & 17.3 \\
\ion{O}{viii}   & 16.003 & 18.0 \\
\ion{Fe}{xix}   & 16.106 & 25.3 \\
\ion{Fe}{xviii} & 15.628 & 26.9 \\
\ion{Fe}{xviii} & 16.002 & 28.6 \\
\ion{Fe}{xviii} & 16.078 & 84.3 \\
\hline
\end{tabular}
\end{center}
\end{table}
We investigated whether these model errors could be caused by the use of an outdated ionisation balance. We fitted the spectra of \src{5044} again using the ionisation balance by \citet{BRY09} instead of the default ionisation balance of SPEX (\citealt{ARN92} for iron and \citealt{ARN85} for the other elements). Statistically, the fit did not improve by this (C-stat 2824 cf. 2418 for the original balance). The fit around the 16\ang line did not improve either. We also recalculated the best fitted model using the forementioned  experimental version of the SPEX line list. This also did not improve the fit near the blend. 
\par Work by \citet{AUD01} using a MEKAL model on \object{Capella}, which has a thermal component with a temperature comparable to our sources, shows that the spectral lines around 16\ang are indeed not fitted well. The fact that the other line blend of \ion{Fe}{xviii} and \ion{Fe}{xix} is correctly fitted in one of the cases hints towards uncertainties in the unresolved spectral lines at these wavelengths. Since most of the lines of iron are fitted correctly when using the abundances that we find, the effect of single poorly fitted lines on the iron abundance should be limited. \citet{WER09} show an indication for resonant scattering of one of the \ion{Fe}{xvii} lines in \src{5813} which could cause a slight underestimation of the iron abundances. Although they try to take this effect into account, the 16\ang line is not perfectly fitted in that work as well. Since the blend at 16\ang also consists of the \ion{O}{viii}$-Ly\alpha$ an resonant scattering effect on this line might also influence the underestimation of the flux at this blend.
\par Line ratios that differ from the ones calculated using the ionisation balance at a given temperature might be caused by non-equilibrium effects. To test for the presence of these effects, the \ac{CIE} model allows for fitting of the ratio between the ionisation balance temperature and the electron temperature. The best-fit ratio we find is  1.21$\pm$0.03 for \src{5044} and 1.30 $\pm$ 0.05 for \src{5813}. The fitting of this parameter improves the C-statistic with $\sim$50. However, the problems with the 14\ang and 16\ang lines are not solved. Moreover, due to the high densities and collisional timescales in the cores of the groups of galaxies studied, it is highly unlikely that non-equilibrium effects are present. 
\par Clusters and groups of galaxies intrinsically contain multiple gas components with different temperatures. Therefore, we need multi-temperature models to constrain temperature structures and abundances correctly \citep[e.g.][]{PET03,KAA04_art}. Knowing this, it is remarkable that \src{5813} shows a single-temperature spectrum in \ac{RGS}. This is fully consistent between the \textit{wdem} and \textit{gdem} fits of the \ac{RGS} data and was previously found by \citet{WER09}. In the cross-dispersion direction of \ac{RGS} we do measure a significant but relatively small rise of the temperature near the outermost bin (between \amin[1] and \amin[2]). In the \ac{EPIC} data, the \textit{gdem} models have a $\sigma_{\mathrm{T}}$ of 0.057$\pm$0.004, corresponding to a \kev[0.09] spread in the temperature.
\par The analysis of \Chandra data by \citet{RAN11} shows a high-resolution temperature map of the inner region of \src{5813} (r$<$\amin[3]). The temperatures range from \kev[$\sim$0.6] to \kev[$\sim$0.7]. To investigate whether some, maybe unresolved, temperature structure would cause an effect comparable to our results, we simulated a spectrum consisting of two \ac{CIE} components.  We choose to set one of these components to have a temperature of \kev[0.6] and put the other at \kev[0.7]. The normalisations were fixed to half the value found for \src{5813} and all other parameters were fixed to the best-fit values derived from the \ac{RGS} data of \src{5813}. We then fitted this spectrum under the assumption of one single \textit{gdem} distribution and found a $\sigma_{\mathrm{T}}$ of 0 and a value for the ion to electron temperature ratio that is the same as we find in our fits. A fit to the \src{5813} data using two distinct \ac{CIE} models caused both temperatures to converge to the value found using a single \textit{gdem} model. The effect found by leaving the ratio between the electron and ionisation balance temperature free could in both sources be caused by the part of the temperature structure that can not be resolved using the models.

\subsection{Abundances}
The oxygen abundances as measured by \ac{EPIC} are lower than what is measured using \ac{RGS}. Oxygen lines can not be well resolved by \ac{EPIC} due to its relatively poor spectral resolution and uncertain calibration \citep[e.g.][]{PLA07}. Since we ignore \pn in the region of the oxygen line, instrumental effects are probably due to the cross-calibration between \ac{RGS} and \mos. Since the \ac{RGS} spectra resolve the \ion{O}{viii} line at $\sim 19$\ang, we use the \ac{RGS} value. For \src{5044} we could fit the \ac{EPIC} spectrum leaving the instrumental power-law normalisation to vary within a factor 2. For \src{5813}, we needed to increase the factor to 4 and to leave the photon index break free. This effect could be due to a possible low contribution of quiescent soft-protons. We compared the abundances and temperature for those fits to a fit leaving only the normalisation free within a factor of 2. The abundances of nickel, argon and sulfur did significantly differ between those fits. We therefore decided not to use the abundances of nickel, argon and sulfur for further analysis in \src{5813}.
\par To check for a potential residual low-mass X-ray binary contribution to our spectra, we added a power-law to the spectral model in \ac{RGS}. This did not improve the fit in any way, whether or not we fixed the photon index to a value of 1.5. For \src{5813} this is in agreement with the results of \citet{RAN11}.

\begin{table}[t]
\caption{Literature values of the parameters for NGC 5044} \label{tab:lit5044}
\centerline{
\begin{tabular}{llll}
\hline
Parameter & T03 & B03 & This work\\
\hline
\hline
Instrument & RGS & ACIS\tablefootmark{a} \& EPIC & RGS \& EPIC\\
$r_{\mathrm{extr}}$ & \amin[1]\tablefootmark{b} & \amin[5] & \amin[2.5]\tablefootmark{b} \& \amin[3]\\
Plasma code & 2 * \textit{CIE} & 2 * \textit{APEC} & \textit{gdem}\\
$kT$ & 1.07(3) & 1.38(2)  & 0.910(5)\\
& 0.804(4) & 0.665(8) & \\
N/Fe & 0.6(2) & --         & 1.5(2) \\
O/Fe & 0.56(2) & 0.42(4)   & 0.60(3) \\
Ne/Fe& 0.56(11) & 0.75(19) & 0.37(10) \\
Mg/Fe& 0.85(5) & 0.87(3)   & 0.75(5) \\
Si/Fe& 0.97(7) & 0.81(2)   & 0.788(12)\\
S/Fe & -- & 0.516(19)      & 0.843(16)\\
Fe   & 0.74(3) & 0.99(4)   & 0.616(16)\\
\hline
\end{tabular}
}
\tablefoot{T03: \citet{TAM03}, B03: \citet{BUO03_II}. Abundance ratios in proto-Solar units by \citet{LOD03}. \tablefoottext{a}{on board of the \Chandra satellite} \tablefoottext{b}{Half width of the region in cross-dispersion direction used for the analysis}}
\end{table}

The values found in the literature for \src{5044} and \src{5813} are summarised in Tables \ref{tab:lit5044} and \ref{tab:lit5813}. Most other work has been carried out using spectra based on different extraction regions, instruments or models. All these can have an effect on the fitted temperatures and absolute abundance values. Abundance ratios do however generally not change with radius \citep[e.g.][]{SIM09}. Because of this, the derived values for those should be comparable. Differences in those values therefore are an indication of the systematic effects introduced by the usage of different instruments and models.

\begin{table}[t]
\caption{Literature values of the parameters for NGC 5813.} \label{tab:lit5813}
\centerline{
\begin{tabular}{lllll}
\hline
Parameter & W09 & R11 & R11 & This work \\
\hline
\hline
Instrument & RGS & RGS & ACIS\tablefootmark{a} &  RGS \& EPIC\\
$r_{\mathrm{extr}}$ & \amin[0.25]\tablefootmark{b} & \amin[0.5]\tablefootmark{b} & \amin[2.9] & \amin[2.5]\tablefootmark{b} \& \amin[3]\\
Plasma code & \textit{CIE} \tablefootmark{c} & 2 * \textit{CIE}\tablefootmark{d} & 2 * \textit{APEC} & \textit{gdem} \\
$kT$ & 0.645(8) & $>$0.4  & 0.671(11) & 0.665(2)\\
& & $<$0.4& 0.35(3) \\
N/Fe & 2.6(1.1) & --         & --       & 1.8(3)   \\
O/Fe & 0.71(15) & 0.80(14)   & 0.27(6)  & 0.73(4)  \\
Ne/Fe& 0.4(3)   & 0.72(12)   & 1.14(12) & 0.58(10) \\
Mg/Fe& --       & --         & 0.83(7)  & 0.86(6)  \\
Si/Fe& --       & --         & 0.95(7)  & 0.94(2)  \\
S/Fe & --       & --         & 1.17(12) & 1.13(5)  \\
Fe   & 0.75(9)  & 0.81(8)    & 0.71(4)  & 0.538(17)\\
\hline
\end{tabular}
}
\tablefoot{W09: \citet{WER09} , R11: \citet{RAN11}. Abundance ratios in proto-Solar units by \citet{LOD03}.\tablefoottext{a}{on board of the \Chandra satellite} \tablefoottext{b}{Half width of the region in cross-dispersion direction used for the analysis} \tablefoottext{c}{Model including four Gaussian components to correct for possibly present microturbulence effects.} \tablefoottext{d}{Spectra were fitted using two \textit{CIE} cooling flow models} }
\end{table}
\par \citet{TAM03} (T03) fitted the \ac{RGS} spectrum of \src{5044} (obsID 0037950101) in the \amin[0]-\amin[1] cross-dispersion region using a model consisting of two temperature components. \citet{BUO03_II} (B03) used a two-temperature \textit{APEC} model to constrain the parameters of \src{5044} using \Chandra. The abundance ratio of oxygen over iron is comparable between the two \ac{RGS} observations and lower in \Chandra. The nitrogen to iron abundance ratio we measure is higher than derived by T03. This may be caused by the different background subtraction, extraction regions and temperature structure model (note that the background is relatively large at this wavelength and therefore sensitive to systematic effects). The same holds for the silicon value derived from \ac{RGS} by T03. The value we derive from \ac{EPIC} does however agree with the value of B03 derived from ACIS and \ac{EPIC}. For the magnesium over iron ratio, the value we measure is somewhat lower than found by both T03 and B03, but still acceptably close. The lower sulfur over iron ratio found by B03 can be explained by their higher temperature, at which the emission in the sulfur lines is higher and hence a lower abundances is needed to obtain the same line power. For neon, the values reported by the three studies do not agree well. Since the neon lines in the spectrum are relatively weak and blend with iron, some systematic effects could still play a role because the often imperfect fit to the iron-L complex. We decided not to use it for further analysis. 
\par \citet{WER09} (W09) fitted the \ac{RGS} spectrum of \src{5813} in a strip with width \amin[0.5] (based on obsID 0302460101) in the cross-dispersion direction. For their fits, they take resonant scattering of the strongest \ion{Fe}{xvii} lines into account. \citet{RAN11} (R11) present results of an \ac{RGS} fit to the same data set, using a strip of \amin[1] and they also present the results of \Chandra analysis. The temperature found by W09 is close to what we measure in the inner bins of the cross-dispersion profile. The oxygen over iron abundance ratio that we find is consistent with W09 and the \ac{RGS} value found by R11. The value reported by R11 based on ACIS is much lower. As we noted before, \ac{RGS} has the resolving power to constrain the oxygen abundance using the \ion{O}{viii}$-Ly\alpha$ line as opposed to the CCD instruments where this line is part of a blend. The nitrogen over iron abundance ratio we find is consistent with the value by W09. There is a good agreement between the \Chandra ACIS results and our work based on \ac{EPIC} for the abundance ratios of magnesium, silicon and sulfur over iron. However, the neon abundance is rather different and it is excluded from further analysis.

\begin{table}[t]
\caption{Abundances reported in \Mes by \citet{WER06}.} \label{tab:Mes}
\centerline{
\begin{tabular}{ll}
\hline
Parameter & value \\
\hline
\hline
$kT_{\text{mean}}$ & 1.82(4) \\
C/Fe & 0.74(13) \\
N/Fe & 1.6(2) \\
O/Fe & 0.59(4) \\
Ne/Fe & 1.25(12) \\ 
Mg/Fe & 0.60(6) \\
Fe & 1.06(3) \\
\hline
\end{tabular}
}
\tablefoot{Abundance ratios in proto-Solar units by \citet{LOD03}. Based on fits to the data using a \textit{wdem} model.}
\end{table}

\par We can compare the abundances that we find to those in other giant ellipticals in the cores of groups or clusters. The central galaxy of the \object{Virgo cluster}, \object{M87}, was studied by \citet{WER06}. We show the values in Table \ref{tab:Mes} and compare our results with their findings because all three sources contain a giant elliptical galaxy. The absolute iron abundance that we find is about half the iron abundance in \Mes. As mentioned before, absolute abundances are however sensitive to bias. The O/Fe ratio found for \src{5044} is close to that of \Mes, while \src{5813} has a somewhat higher value. The N/Fe ratio found for our groups is in good agreement with \Mes. We do find a value of C/Fe which is about a factor of 2 higher than in \Mes. It should however be noted that the model background of the \ac{RGS} suffers from severe problems in the CCD chip on which the carbon line is projected. 

\par \citet{WER09} present the \ac{RGS}-measured abundances of five bright elliptical galaxies, including \src{5813}. The fits are based on single-temperature models and include corrections for resonant scattering effects. The abundances are shown in Table \ref{tab:resgals}.
\begin{table}[t]
\caption{Abundances reported by \citet{WER09}.} \label{tab:resgals}
\centerline{
\begin{tabular}{lllll}
\hline
Parameter & \object{NGC 4636} & \object{NGC 1404} & \object{NGC 4649} & \object{NGC 4472} \\
\hline
\hline
$kT$ & 0.606(6) & 0.608(9) & 0.774(7) & 0.781(6) \\
N/Fe & 2.5(6) & 3.4(1.3) & 1.5(9) & 1.6(6) \\
O/Fe & 0.85(11) & 0.87(18) &  0.7(2) & 0.64(10) \\
Ne/Fe& 0.60(16) & 1.2(4) &  1.5(5) & 1.4(3) \\
Fe   & 0.52(3) & 0.67(8) & 0.87(18) & 0.83(8) \\	
\hline
\end{tabular}
}
\tablefoot{Abundance ratios in proto-Solar units by \citet{LOD03}.}
\end{table}
The fits presented in this table are based on relatively short observations (17 -- 81 ks) which translates in relatively large error bars on the parameters. The ratio of nitrogen to iron is higher than solar for all objects. It appears that data from the shorter observations hint towards higher values for this ratio although we do not have enough data to confirm this. For \src{5813} the longer observation shows a lower abundance ratio. For the oxygen over iron ratio, the values of the abundances are all comparable with what we measure in our groups. As mentioned before, the neon abundance is very hard to constrain. The measured values are sometimes hardly significant and there is some spread. The absolute iron abundances do also show some spread. Those are however, as mentioned before, sensitive to biases.
\par The abundances of carbon, nitrogen, oxygen (in \ac{RGS}), silicon and sulfur can be constrained based on single, isolated lines and are much more robust than the abundances of neon and magnesium because these metals are only constrained using lines that are blended with iron lines. The calcium and argon abundances measured are high with large error bars. Moreover they may be somewhat biased by residual background in the relevant energy band. The peak of the calcium and argon lines is $5-10\%$ above the continuum, while the continuum shows systematic offsets of $10-20\%$ at the wavelengths of these elements (see e.g. Fig. \ref{fig:5813-mpspec}). In \src{5044}, the relative abundances measured with \ac{RGS} are also consistent for the two \ac{DEM} parametrisations. The iron abundance itself differs significantly, as well as the normalisation of the hot gas model.
\par The abundance ratios give an insight in the processes that create the metals in a cluster or group of galaxies. Most of the metals in galaxies are created by supernovae. Some metals, especially carbon and nitrogen, are however not so efficiently created by supernovae and result from stellar winds of main sequence or AGB stars. We discuss both points in a bit more detail. 
\subsection{Relative supernova yields}
For years, the oxygen and iron abundances of our Galaxy have been used to determine the ratio between type Ia and core-collapse supernovae \citep[see e.g. the reviews by][]{RAN91,WOO02}. In general type Ia supernovae are believed to yield most of the iron in the Galaxy while core-collapse supernovae would be responsible for the oxygen. The spread of the yields between the different models is however large.  Attempts have been made to constrain the supernova ratios and models using the abundances in the hot gas of clusters of galaxies \citep[see e.g.][]{PLA07}. 
\par The abundances that we measure in the gas do not directly reflect the chemical yield of supernovae during the total history of the group. But the abundances do provide constraints on the ratio of type Ia to core-collapse supernovae contributing to the gas \citep[see][]{HUM06,PLA07}. If we assume that all iron and oxygen has been created by supernovae, we can use the ratio of these two abundances to find the ratio between type Ia and core-collapse supernovae that enriched the gas. We write
\begin{equation}
 \left(\frac{\mathrm{X}}{\mathrm{Fe}}\right)_{\mathrm{group}} =  f\left(\frac{\mathrm{X}}{\mathrm{Fe}}\right)_{\mathrm{SNIa}} + (1-f)\left(\frac{\mathrm{X}}{\mathrm{Fe}}\right)_{\mathrm{SNcc}} \label{eq:ffac}
\end{equation}
\acused{IMF}
Here $\mathrm{X}$ is the abundance of an element, $\mathrm{Fe}$ the iron abundance and $f$ represents the fraction of the iron that was created by type Ia supernovae. The supernova models that we use are the best-fit models found by \citet{PLA07}. For core-collapse supernovae, this is the model by \citet{NOM06}, using stars with $Z=0.02$ and a Salpeter \acl{IMF} \citep[\acs{IMF}; ][]{SAL55}. 
\par Since the yield of a core-collapse supernova depends on the mass, the typical core-collapse yield we use is the \ac{IMF} weighted integrated yield \citep{TSU95}.
\par For the Type Ia supernovae, we use four different models. First we use two delayed detonation (DDT) models by \citet{BAD03}. The DDTc model compared best to the abundances found in the \object{Tycho} supernova remnant \citep{BAD06}. \citet{BAD08} propose that \object{SNR 0509$-$67.5} is best fitted by the DDTa model from \citet{BAD03}. The difference between the DDT models is due to the different input parameters for the density of the ambient medium and the kinetic energy of the ejecta. The DDTa model predicts lower amounts of silicon to calcium than DDTc (Table \ref{tab:SNcomp}). We also compare our data to the W7 and WDD2 model from \citet{IWA99} because they are commonly used in the literature. These two models do also predict higher amounts of silicon and sulfur than the DDTa model. There is also some spread in the yields of argon and nickel between the four models. The DDTc model predicts the highest argon value, which is still lower than we measure for \src{5044}. The nickel abundance predicted by the W7 model is higher than what we find while for the other models it is significantly lower. The amount of produced iron predicted by the DDTa model is higher than in the other models. This translates in the lower number ratio for comparable $f$.
\par To find the value for $f$, we use the oxygen over iron abundance. We then use this value for $f$ to predict the abundance ratios of the other elements expected by the models and compare these to our results. This is shown in Table \ref{tab:SNcomp}. 
\begin{table}[t]
\caption{Comparison of the Supernova yields to the measured abundances.} \label{tab:SNcomp}
\centerline{
 \begin{tabular}{llllll}
\hline
Parameter & DDTc & DDTa & WDD2 & W7 & measured \\
\hline \hline
\multicolumn{6}{c}{NGC 5044} \\
\hline
$f$ & 0.84 & 0.83 & 0.83 & 0.84 & \\
$\frac{N_{Ia}}{N_{Ia}+N_{cc}}$ & 38\% & 31\% & 36\% & 39\% & \\
C / Fe & 0.19 & 0.19 & 0.19 & 0.21 & 1.5 $\pm$ 0.3\\
N / Fe & 0.22 & 0.23 & 0.22 & 0.22 & 1.5 $ \pm$ 0.2\\
O / Fe & 0.6  & 0.6 & 0.6 & 0.6 & 0.60 $\pm$ 0.03 \\
Mg / Fe & 0.47 & 0.48 & 0.49 & 0.48 & 0.73 $\pm$ 0.05\\
Si / Fe & 0.67 & 0.50 & 0.74 & 0.66 & 0.788 $\pm$ 0.012 \\
S / Fe  & 0.76 & 0.51 & 0.74 & 0.62 &  0.843 $\pm$ 0.016\\
Ar / Fe & 0.58 & 0.36 & 0.49 & 0.38 & 0.789 $\pm$ 0.013\\
Ca / Fe & 0.96 & 0.56 & 0.71 & 0.47 & 1.86 $\pm$ 0.13 \\
Ni / Fe & 1.01 & 1.11 & 1.16 & 2.48 & 1.60 $\pm$ 0.05\\
\hline
\multicolumn{6}{c}{NGC 5813} \\
\hline
$f$ & 0.80 & 0.79 & 0.80 & 0.80 & \\
$\frac{N_{Ia}}{N_{Ia}+N_{cc}}$ & 32\% & 25\% & 32\% & 33\% & \\
C / Fe & 0.23 & 0.23 & 0.24 & 0.25 & 1.7 $\pm$ 0.4\\
N / Fe & 0.27 & 0.27 & 0.28 & 0.27 & 1.8 $\pm$ 0.3\\
O / Fe & 0.73 & 0.73 & 0.73 & 0.73 & 0.73 $\pm$ 0.04\\
Mg / Fe & 0.57 & 0.57 & 0.59 & 0.59 & 0.86 $\pm$ 0.06\\
Si / Fe & 0.73 & 0.57 & 0.81 & 0.73 & 0.99 $\pm$ 0.03\\
Ca / Fe & 0.98 & 0.60 & 0.74 & 0.51 & 1.9 $\pm$ 0.5\\
\hline
\multicolumn{6}{c}{\Mes\tablefootmark{a}} \\
\hline
$f$ & 0.84 & 0.83 & 0.83 & 0.84 & \\
$\frac{N_{Ia}}{N_{Ia}+N_{cc}}$ & 38\% & 31\% & 36\% & 39\% & \\
C / Fe & 0.19 & 0.19 & 0.19 & 0.21 & 0.74 $\pm$ 0.13\\
N / Fe & 0.22 & 0.23 & 0.22 & 0.22 & 1.62 $\pm$ 0.21 \\
O / Fe & 0.6  & 0.6 & 0.6 & 0.6 & 0.60 $\pm$ 0.03 \\
Ne / Fe & 0.92 & 0.95 & 0.93 & 0.91 & 1.25 $\pm$ 0.12\\
Mg / Fe & 0.47 & 0.48 & 0.49 & 0.48 & 0.60 $\pm$ 0.06\\
\hline
\end{tabular}
}
\tablefoot{ Yields based on the models by \citet{BAD03}, \citet{IWA99} and \citet{NOM06}. Abundance ratios in proto-Solar units by \citet{LOD03}. The ratio of the iron created by \ac{SN}Ia to the total amount of iron (see Eq. \ref{eq:ffac}), $f$, is calculated based on the O/Fe ratio using the assumption that all oxygen and iron has been created by supernovae. \tablefoottext{a}{Values measured by \citet{WER06} are compared with the models used in this work.}}
\end{table}
We find a value for $f$ which is consistent with the value found for \src{5044} by \citet{BOH05}, based on the abundances reported by \citet{BUO03_II}, using oxygen, iron and silicon abundances. The number is also consistent with \Mes \citep{WER06}. This means that the the stellar population of our sources is similar to the population of \Mes. If we use the abundances of the solar neighbourhood as input (i.e. O/Fe$\equiv$1), we find a ratio of $\sim$20\% of the supernovae being of type Ia. The stellar population of the group is significantly older than in the surroundings of our Sun. By fitting supernova yields to multiple abundance values, \citet{PLA07} derived from a sample of clusters that 44$\pm$5\% of the contributing supernovae is of type Ia. This means that the stellar population in our groups seems to be slightly younger than or comparable to the average population in clusters of galaxies. The magnesium abundance predicted by the supernova models is lower than the value we measured. The measured silicon abundances are higher than the values found using the DDTc model, and even more higher than the DDTa model. The DDTc model reproduces a high calcium abundance, but this is still lower than our measurements.
\par We compare the abundances with the metallicities of stars in the disk of our own Galaxy \citep[by][]{CHE00,SHI02,BEN06}. Based on this we can indicate what the differences are between the large ellipticals in the centre of groups and our Galactic environment. For comparison with Galactic stars, we use $[X/Y] = \log(X/Y) - \log(X/Y)_{\odot}$ as parameter, converting the abundances from the proto-Solar abundances of \citet{LOD03} to the photospheric solar abundances of \citet{LOD03}. However, the difference between the photospheric and solar abundance is 0.07 dex for all elements for which we present a ratio (C, N, O, Fe). The absolute values of the abundances change, but the ratios are the same in both abundance systems. The abundance ratios measured in Galactic stars are calibrated on a measured Solar spectrum. We assume that this difference between the measured Solar spectrum and the recent Lodders abundances are small compared to our uncertainties.
\par Since star formation has continued in our Galaxy, we can assume that the metallicities in young stars, which are located in the disk of the Galaxy, are comparable to the metallicities in the ambient medium. The oxygen to iron abundance ratio which we find for our sources is [O/Fe] = $-$0.222 $\pm$ 0.022 at [Fe/H] = $-$0.210 $\pm$ 0.011 for \src{5044} and [O/Fe] = $-$0.137 $\pm$ 0.018 at [Fe/H] = $-$0.269 $\pm$ 0.014 for \src{5813}. The ratios are low compared to the set of stars by \citet{BEN06}. Therefore we can state that the contribution of core-collapse supernovae is relatively low in our sources compared to the core-collapse contribution in our Galaxy.
\subsection{Carbon and  nitrogen progenitors}
\par For nitrogen and carbon, the measured abundance values are respectively factors 8 and 7 higher than predicted by supernovae only. For nitrogen, this number is much lower than the factor of 313 which was found by \citet{WER06}. The difference mostly rises from the fact that the supernova models we used for type Ia's include nitrogen creation. For the same reason the ratio we find for carbon is lower than the value of 10 reported by \citet{WER06}.
\par Conclusions about the parent stellar population based on the carbon and nitrogen abundance in the hot intra-group medium  should be drawn with care. First of all, these metals are still being created and a fraction could still be locked up in stars. On the other hand, the stellar mass fraction of groups and clusters is 10--20\% of the gas mass \citep{ETT09}. This amount is comparable to the corresponding uncertainties on carbon and nitrogen and therefore the effect of those metals being locked up in stars is limited. Furthermore, the secondary creation of carbon and nitrogen depends on the metallicity of the environment in which a star was formed. The formation of the less massive stars, which at this moment still enrich the gas, did however happen long ago during a relatively short period of the groups' life time. The metallicities of those stars around the time they formed are likely to be comparable to each other. In addition, our observations do not directly show how and when elements were ejected into the hot medium. As mentioned before, multiple mechanisms exist that enrich the medium with metals created in the galaxies. All mechanisms act on their typical time scale. Therefore, the measured abundance ratios are the ratios in the intra-group gas, and not necessarily the ratios in the constituent galaxies. Since core-collapse supernovae, which can eject metals in a wind, seem to have decreased significantly in rate during the life time of the group, it is likely that other enrichment mechanisms took over. It was shown that jets from the \ac{AGN} in the centre of the giant elliptical galaxy can drag out metals from the galaxies into the hot gas \citep{SIM08,SIM09,KIR11}. Both \src{5044} and \src{5813} show evidence of \ac{AGN} activity \citep{DAV09,RAN11}, therefore \ac{AGN} uplift is probably the currently dominating enrichment mechanism.
\begin{table}[t]
\caption{Comparison of the measured abundance ratios to the stars in the disk of our galaxy.} \label{tab:Galstars}
\centerline{
\begin{tabular}{llll}
\hline
Ratio & \src{5044} & \src{5813} & Typical range in disk stars \\
\hline
\hline
$[$N/Fe$]$ & 0.18$\pm$0.06 & 0.26$\pm$0.07 & -0.2$\leq$ $[$N/Fe$]$ $\leq$0.2\tablefootmark{a} \\
$[$N/O$]$  & 0.40$\pm$0.06 & 0.39$\pm$0.08 & -0.5$\leq$ $[$N/O$]$ $\leq$0.15\tablefootmark{a} \\
$[$C/Fe$]$ & 0.18$\pm$0.09 & 0.23$\pm$0.10 & -0.2$\leq$ $[$C/Fe$]$ $\leq$0.2\tablefootmark{b}\\
\hline
\end{tabular}
}
\tablefoot{Abundance ratios in solar units. \tablefoottext{a}{\citet{CHE00} and \citet{SHI02}.} \tablefoottext{b}{From both the thick and thin disk stars of \citet{BEN06}.}
}
\end{table}
\par The exact scenario for the creation of nitrogen is not yet fully understood \citep[see e.g.][]{SCH07, PIP09}. Models suggest that core-collapse supernovae produce much more oxygen than nitrogen. Nitrogen may instead be produced by low and intermediate mass stars. A high nitrogen over oxygen ratio would then hint towards lower present star formation. The comparison of the nitrogen to oxygen ratios of the two studied groups and the ranges of those values quoted in the literature \citep{CHE00,SHI02,BEN06} are shown in Table \ref{tab:Galstars}. Based on the high nitrogen to oxygen abundance ratio in the two studied groups, we can conclude that following the initial starburst the star-formation rate in these giant elliptical galaxies was significantly lower than in the Milky Way. This agrees well with the general assumption that the present star formation rate in elliptical galaxies is negligible. Due to the continuous evolution of low and intermediate mass stars, nitrogen is still being enhanced while hardly any new oxygen is created.
\par When we compare the carbon and nitrogen to iron abundance ratios  that we find to the typical abundance ratios in disk stars of our own Galaxy, we conclude that both the carbon and nitrogen abundances are comparable to the stars with the highest ratios in disk of our Galaxy. The fact that the nitrogen and carbon abundance ratios are higher than the ones in most of the stars of our Galaxy, while the abundance ratio of oxygen over iron is relatively low seems to point towards intermediate mass stars in \ac{AGB} phase as sources of carbon and nitrogen creation in our sources, because the mass range of the progenitors of white dwarfs  involved in type Ia supernovae (3 -- 8 $M_{\odot}$) is the same as for \ac{AGB} stars. It should be noted that this conclusion is based on simple arguments and should be verified by more detailed population synthesis studies.

\section{Conclusions}
We present the results of the analysis of \src{5044} and \src{5813} data, obtained with the \ac{RGS} and \ac{EPIC} instruments aboard \xmm. From the analysis we conclude that:
\begin{itemize}
\item The relatively low ratio of the oxygen over iron abundance, combined with the high abundance ratios to iron for carbon and nitrogen suggest that the bulk of the carbon and nitrogen are not created by massive stars which end their lives as core-collapse supernovae.
\item From the oxygen and iron abundance, and under the assumption of a Salpeter \ac{IMF}, we find for both studied groups that about 80\% of the iron is formed by type Ia supernovae. This translates to as roughly 30--40\% of supernovae contributing being of type Ia while the other $\sim$ 60--70\% of supernovae then are core collapse.
\item The relative contribution of core collapse supernovae to type Ia supernovae is comparable with what was previously found for \Mes. This means that the giant elliptical galaxies in our study have a stellar and chemical evolution history comparable to the central giant elliptical galaxy of the virgo cluster. Both stellar populations tend to be older than in the surroundings of our Sun.
\item A \textit{wdem} and \textit{gdem} model give comparable results in terms of abundances for the \ac{RGS} observations of \src{5044} and \src{5813}
\item Low temperature plasma models have problems fitting the line blends of \ion{Fe}{xviii}, \ion{Fe}{xix} and \ion{O}{viii} near 14\ang and 16\ang.
\end{itemize}
\begin{acknowledgements}
The anonymous referee is thanked for the useful comments on how to improve the science of the paper. We thank Joke Claeys, Sjors Broersen and Ton Raassen for their useful comments on the manuscript. Carles Badenes is kindly thanked for providing us the yields of his supernova models. The Netherlands Institute for Space Research (SRON) is supported financially by NWO, the Netherlands Organisation for Scientific Research. Based on observations obtained with XMM-Newton, an ESA science mission with instruments and contributions directly funded by ESA Member States and NASA.
\end{acknowledgements}
\bibliographystyle{aa}
\bibliography{biblio}

\end{document}